\documentclass[twocolumn]{aastex61}

\received{}
\revised{}
\accepted{}

\submitjournal{The Astrophysical Journal}

\usepackage{rotating}

\shorttitle{Merger Shocks}
\shortauthors{Ha et al.}

\begin{document}

\title{Properties of Merger Shocks in Merging Galaxy Clusters}

\author{Ji-Hoon Ha}
\affil{Department of Physics, School of Natural Sciences UNIST, Ulsan 44919, Korea}
\author{Dongsu Ryu}
\affiliation{Department of Physics, School of Natural Sciences UNIST, Ulsan 44919, Korea}
\author{Hyesung Kang}
\affiliation{Department of Earth Sciences, Pusan National University, Busan 46241, Korea}
\correspondingauthor{Dongsu Ryu}
\email{ryu@sirius.unist.ac.kr}

\begin{abstract}
 
% limit 250 words
% currently 250 words

{ X-ray shocks and radio relics detected in the cluster outskirts are commonly interpreted as shocks induced by mergers of sub-clumps. We study the properties of merger shocks in merging galaxy clusters, using a set of cosmological simulations for the large-scale structure formation of the universe. As a representative case, we here focus on the simulated clusters that undergo almost head-on collisions with mass ratio $\sim2$. 
Due to the turbulent nature of the intracluster medium, shock surfaces are not smooth, but composed of shocks with different Mach numbers. As the merger shocks expand outward from the core to the outskirts, the average Mach number, $\left<M_s\right>$, increases in time. We suggest that the shocks propagating along the merger axis could be manifested as X-ray shocks and/or radio relics. The kinetic energy through the shocks, $F_\phi$, peaks at $\sim1$ Gyr after their initial launching, or at $\sim1-2$ Mpc from the core. Because of the Mach number dependent model adopted here for the cosmic ray (CR) acceleration efficiency, their CR-energy-weighted Mach number is higher with $\left< M_s \right>_{\rm CR}\sim3-4$, compared to the kinetic-energy-weighted Mach number, $\left<M_s\right>_{\phi}\sim2-3$. Most energetic shocks are to be found ahead of the lighter dark matter (DM) clump, while the heavier DM clump is located in the opposite side of clusters. Although our study is limited to the merger case considered, the results such as the means and variations of shock properties and their time evolution could be compared with the observed characteristics of merger shocks, constraining interpretations of relevant observations.}

\end{abstract}


\section{Introduction} 
\label{sec:s1}

The current $\Lambda$CDM cosmology favors the hierarchical structure formation, where small clumps first formed and continuously merged to become galaxy clusters. Shock waves are naturally induced in the intracluster medium (ICM) during the hierarchical structure formation. Since the gas in clusters is in the form of hot tenuous plasma, these shocks are collisionless as in other astrophysical environments. They heat the gas and, at the same time, accelerate cosmic rays (CRs) via diffusive shock acceleration (DSA) \citep[see, e.g.,][]{bell1978, blandford1978, drury1983}. Using cosmological hydrodynamic simulations for the large-scale structure (LSS) formation of the universe, the properties and roles of shocks in the ICM as well as around clusters have been extensively studied \citep{miniati2000, ryu2003, pfrommer2006, kang2007, skillman2008, hoeft2008, vazza2009, schaal2015}.

Such studies have shown that {\it external accretion shocks} form around clusters, when the void gas of $T \sim 10^4$ K accretes onto them. With the accretion velocity of $\sim 10^3\ {\rm km\ s^{-1}}$ and the sound speed of $c_s \sim 10\ {\rm km\ s^{-1}}$, their Mach number is very high, of order $M_s \sim 100$. Yet, due to the low density, the kinetic energy flux through the shock surface, $f_\phi = (1/2) \rho_1 v_s^3$, is small (where $\rho_1$ is the preshock gas density and $v_s$ is the shock speed), and hence external shocks are not energetically important. Inside clusters, {\it internal shocks} are induced by mergers of clumps and infall of the warm-hot intergalactic medium (WHIM) along filaments, as well as turbulent flow motions \citep[see, e.g.,][]{ryu2003}. They form in the hot ICM that were heated by previous shock passages, and hence have lower Mach numbers. But due to the higher gas density, internal shocks have larger $f_\phi$, and thus play more important roles in heating ICMs and producing CRs, compared to external shocks
.

Among internal shocks, {\it turbulent shocks}, produced by turbulent flow motions, are mostly weak with at most $M_s \lesssim 2$ \citep{porter2015}, since ICM flow motions are subsonic with turbulent Mach number $M_t \sim 0.5$ \citep[see, e.g.,][]{ryu2008, vazza2011a, miniati2014, brunetti2014, vazza2017}. {\it Infall shocks} are formed by continuous infall of the WHIM of $T \sim 10^5 - 10^7$ K, often with streams of minor mergers, into the hot ICM of $T \sim 10^7 - 10^8$ K \citep[see, e.g.,][for observations of infall shocks]{brown2011, pfrommer2011, ogrean2013a}. They can have higher Mach numbers of up to $M_s \sim 10$ \citep{hong2014}. With relatively high Mach numbers, infall shocks could be sites for efficient CR acceleration. But they form mostly in the cluster outskirts, since the gas accretion from filaments normally halts around the virial radius and does not penetrate into the core. Moreover, they have small cross sections.

The shocks induced as a consequence of ``major mergers'' are called {\it merger shocks}. A merger of $M \sim 10^{13} M_\sun$ clumps with speed $\sim 10^3\ {\rm km\ s^{-1}}$ involves the kinetic energy of $E_{\rm merg} \sim 10^{62}$ ergs, and a substantial fraction of it is dissipated at merger shocks in the time-scale of $\sim 1 - 2$ Gyr \citep[see, e.g.][]{markevitch2007}. Such merger shocks should be energetic enough to result in observable phenomena; so most shocks observed in X-ray and/or radio, usually in the outskirts of merging clusters, are interpreted as merger shocks.

The best known example of merger shock from X-ray observation is the one in the so-called Bullet Cluster (1E 0657-56) \citep{markevitch2002}. So far dozens of shocks have been observed with {\it Chandra}, {\it XMM-Newton}, and {\it Suzaku} \citep[see, e.g.,][]{markevitch2005, ogrean14, itahana2015, dasadia2016}. They are found typically at distance $d_s \gtrsim 1$ Mpc from the cluster center, and mostly weak with $M_s \sim 1.5 - 3$.

``Radio relics'' are known to be the radio manifestation of ICM shocks. Their emission is interpreted as synchrotron radiation from CR electrons accelerated at shocks associated with them. Well studied radio relics include the so-called Sausage relic in the cluster CIZA J2242.8+5301 \citep[e.g.,][]{vanweeren2010, stroe13}, double relics in ZwCl 0008.8+5215 \citep[e.g.,][]{vanweeren2011b} and PLCK G287.0+32.9 \citep[e.g.,][]{bagchi11, bonafede2014}, and the so-called Toothbrush relic in RX J0603.3+4214 \citep[e.g.,][]{vanweeren2012, vanweeren2016}. In addition, about a hundred radio relics have been observed so far \citep[see, e.g.,][for reviews]{feretti12, brug12a, brunetti2014}. They are also found at $d_s \gtrsim 1$ Mpc, but the associated shocks are on average stronger than X-ray shocks, with $M_s$ as high as $\sim 4.5$.

A notable point is that the shock parameters inferred from X-ray and radio observations for the same object do not always agree with each other. In the case of the Sausage relic, for instance, the Mach number estimated with the radio spectral index near the edge (shock surface) is $M_{\rm radio} \approx 4.6$ \citep{vanweeren2010}, while the value estimated with the discontinuity in X-ray observations is smaller with $M_{\rm X} \approx 2.5 - 3.1$ \citep{ogrean14, akamatsu15}. In the case of the Toothbrush relic, the radio spectral index indicates $M_{\rm radio} \approx 2.8$ \citep{vanweeren2016}, but X-ray observations reveal $M_{\rm X} \lesssim 2$ \citep{itahana2015, vanweeren2016}.

It was argued that the discrepancy between $M_{\rm X}$ and $M_{\rm radio}$ could be resolved by the reacceleration model in which a shock with $\sim M_{\rm X}$ sweeps through and reaccelerates pre-existing ``fossil CR electrons'' of a flat energy spectrum consistent with the observed radio spectrum \citep[e.g.,][]{kangryu15}. But recently, \citet{kang2017} suggested that, considering short cooling time scales of GeV electrons, it might be unrealistic to generate and/or maintain such flat-spectrum fossil CR electrons, so a shock with $\sim M_{\rm radio}$ is required to reproduce the aforementioned radio observations. On the other hand, from mock X-ray and radio observations of relic shocks in clusters from LSS formation simulations, \citet{hong15} showed that the surfaces of ICM shocks are inhomogeneous with different $M_s$'s at different parts, and X-ray observations pick up the parts with higher shock energy flux but lower $M_s$, while radio emissions come preferentially from the parts with higher $M_s$ and so higher electron acceleration. As a result, $M_s$ inferred from X-ray discontinuities tends to be lower than that from radio spectral indices, explaining the discrepancy of $M_s$ in X-ray and radio observations.

The reacceleration scenario was partly motivated by the scarcity of radio relics. It is expected that most merger shocks would appear as radio relics, yet the fraction of X-ray luminous merging clusters with observed radio relics is order of $\sim 10 \%$ \citep[see, e.g.,][]{feretti12}. In addition, some X-ray shocks do not exhibit radio relics \citep[see, e.g.,][]{russell2011}. In the reacceleration scenario, merger shocks light up as radio relics only when they encounter clouds of fossil electrons left over, for instance, from either radio jets or previous episodes of shock/turbulence accelerations \citep[see, e.g.,][for observations interpreted to reveal the reacceleration scenario]{bonafede2014, shimwell15, vanweeren2017}.

\begin{deluxetable*}{cccccc}[t]
\tablecaption{Merging Cluster Sample \label{tab:t1}}
\tabletypesize{\scriptsize}
\tablecolumns{6}
\tablenum{1}
\tablewidth{0pt}
\tablehead{
\colhead{} &
\colhead{{ $M_{\rm heavy}/M_{\rm light}$ $^a$}} &
\colhead{$T_{X,\rm heavy}$ (keV)$^b$} & 
\colhead{$T_{X,\rm light}$ (keV)$^b$} &
\colhead{$T_X$ (keV)$^c$} &
\colhead{$z_i$ $^d$} 
}
\startdata
Cluster 1 & { 1.84} & 4.26 & 2.99 & 5.12 & 0.36\\
Cluster 2 & { 1.97} & 3.88 & 2.20 & 4.65 & 0.35\\
Cluster 3 & { 1.96} & 4.08 & 2.33 & 4.92 & 0.23\\
Cluster 4 & { 2.00} & 3.79 & 2.15 & 4.55 & 0.30\\
Cluster 5 & { 1.99} & 3.83 & 2.18 & 4.60 & 0.25\\
\enddata
\tablenotetext{a}{  Ratio of total (baryons and DM) virial masses of two merging clumps, estimated at 0.174 Gyr before the axial shock launching time.}
\tablenotetext{b}{X-ray weighted temperatures of two merging clumps, estimated at { 0.174 Gyr} before the axial shock launching time.}
\tablenotetext{c}{X-ray weighted temperature of merged clusters, estimated at 1 Gyr after the axial shock launching time.}
\tablenotetext{d}{  Redshift of axial shock launching time (see Section \ref{sec:s3.2}).}
\end{deluxetable*}

Weak lensing observations have enabled the reconstruction of mass distribution in clusters. The technique has been applied to several merging clusters, imposing constraints on the interpretation and modeling of observed shocks. A weak lensing study of the Bullet Cluster, for instance, found dark matter (DM) clumps behind shocks, whose peaks are offset from the X-ray peaks \citep{clowe2004}. A weak lensing mass reconstruction of CIZA J2242.8+5301 by \citet{jee2015} revealed two DM clumps of almost equal masses, whose distributions are offset from the galaxy distribution as well as the X-ray emission. \citet{okabe2015}, on the other hand, argued that the clump behind the Sausage relic is less massive, and the one in the other side of the cluster and close to the peak of X-ray emission is about twice more massive. {  In addition, mass reconstructions identified, for instance, two DM clumps of mass ratio $\sim 5$ in ZwCl 0008.8+5215 \citep{golovich2017}, two dominant DM clumps of mass ratio $\sim 3$ and a few smaller clumps in RX J0603.3+4214 \citep{jee2016}, and one dominant DM clump and several smaller clumps in PLCK G287.0+32.9 \citep[see][]{finner17}. In these clusters, heavy clumps are located behind the main relics.}

Along with shocks, ``cold fronts'' are commonly observed in merging clusters. Cold fronts refer to the structures with opposite gradients of density and temperature, or contact discontinuities in fluid dynamics. Since early reports in \citet{markevitch2000} and \citet{vikhlinin2001a}, a number of cold fronts have been observed \citep[see, e.g.,][]{markevitch2002, markevitch2007}. They are often modeled as the borders of cool clumps \citep[see, e.g.,][]{bourdin2013}, or in some cases, they are thought to be produced as a result of sloshing motions of clumps \citep[see, e.g.,][]{zuhone2010}. Some of them appear behind merger shocks in merging clusters, typically about a half way from the cluster core to shocks \citep[see, e.g.,][]{markevitch2002, emery2017}. Weak lensing observations indicate that in some cases, their locations are close to the peak of the DM distribution \citep[see, e.g.,][]{clowe2006, okabe2008}.

All the above observations tell us that the nature of merger shocks need to be understood and described in the context of the LSS formation, along with other observables such as X-ray and DM distributions.

Previous studies about shock waves inside and around clusters \citep[see, e.g.,][and the references mentioned above]{ryu2003} mainly concerned on the overall statistics of all cluster shocks, and so did not particularly highlight merging clusters. \citet{paul2011} and \citet{schmidt2017} studied merging clusters in the context of the LSS formation, but did not analyze the properties of merger shocks in detail. \citet{springel2007}, \citet{vanweeren2011a} and \citet{molnar2017a,molnar2017b}, for instance, on the other hand, simulated and studied idealized binary mergers, but in a controlled-box, {  modeling specific objects, i.e., merger shocks in 1E 0657-56, CIZA J2242.8+5301, or ZwCl 0008.8+5215}.

{  In this paper, we study merger shocks  in cosmological environments, reproduced with a set of hydrodynamic simulations for the LSS formation of the universe. As far as we know, this is the first attempt to simulate cluster merger shocks and analyze their properties {\it in the context of the LSS formation}, as opposed to the idealized binary merger simulations cited above. Observed merger events have a wide range of values for clump number, cluster mass, and impact parameter \citep[see, e.g.,][]{clowe2004,okabe2015,jee2016,golovich2017}, and each shows distinctive features. We here focus on major merger events involving ``almost head-on collisions'' (with small impact parameters) of clumps with ``mass ratio $\sim 2$'', because they are most likely to be observed with giant radio relics such as the Sausage relic associated with  CIZA J2242.8+5301 \citep[see, e.g.,][]{okabe2015}. We leave the exploration of mergers with different mass ratios and impact parameters as future works.} By examining the spatial distributions of gas, X-ray emission, and DM, we identify merger-driven shocks and describe their properties. {  Especially, we quantify their properties in the realistic {\it turbulent ICM}, which could be done only with full LSS formation simulations. The quantities, such as the means and variations of $v_s$ and $M_s$ at shock surfaces and their time evolution, should provide inputs for detailed modeling of synchrotron emissions \citep[see, e.g.,][]{kangryu15,kang2017} and also constrain $M_{\rm X}$ and $M_{\rm radio}$ inferred from X-ray and radio observations of radio relics \citep[see, e.g.,][]{hong15}.}

The paper is organized as follows. In Section \ref{sec:s2}, details of numerical simulations and the compilation of sample merging clusters are described. In Section \ref{sec:s3}, the identification of merging shocks is described, and then the properties of merger shocks and their time evolution are presented. A summary follows in Section \ref{sec:s4}.

\section{Numerics}
\label{sec:s2}

\subsection{Simulations and Cluster Sample}
\label{sec:s2.1}

To generate a sample of merging clusters used in this study, we performed numerical simulations of the LSS formation of the universe for a $\Lambda$CDM cosmology model with baryon density $\Omega_{\rm BM}=0.044$, DM density $\Omega_{\rm DM}=0.236$, cosmological constant $\Omega_\Lambda=0.72$, rms density fluctuation $\sigma_8=1.05$, Hubble parameter $h\equiv H_0/(100\ {\rm km}\ {\rm s}^{-1}{\rm Mpc}^{-1})=0.7$, and primordial spectral index $n=0.96$. Except $\sigma_8$, the parameters are consistent with the WMAP7 data \citep[][]{komatsu2011}. {  While $\sigma_8 \approx 0.82$ is the value best fitted to the WMAP7 data, we adopted a slightly larger $\sigma_8$ to enhance the number of massive clusters formed in the simulations. Previously larger $\sigma_8$'s were often used for cluster simulations, arguing that the properties of individual clusters and shock waves there are not very sensitive to $\sigma_8$ \citep[see, e.g.,][]{thomas1998,vazza2009}.}

Simulations were performed using a PM/Eularian hydrodynamic cosmology code \citep[][]{ryu1993}. A cubic box of comoving size of $50h^{-1}$ Mpc with periodic boundaries was employed. A grid of $1024^3$ uniform zones was used, so the spatial resolution is $\Delta l = 48.8h^{-1}$ kpc. Nongravitational effects such as radiative and feedback processes were not included.

{  Sample clusters were compiled from a number of simulations with different realizations of initial condition. As noted in the Introduction, we here focus on mergers with clump mass ratio $\sim 2$, going through almost head-on collisions (specifically, the impact parameter $\lesssim 2 \Delta l \equiv 140$ kpc). In addition, we constrained the epoch of the launching of axial shocks to the redshift range of $0.23 \lesssim z_i \lesssim 0.36$, ensuing the shocks have the best chance to be observed in X-ray and radio at $0.14 \lesssim z \lesssim 0.25$ ($\sim 1$ Gyr after the shock launching, see Section \ref{sec:s3.2}). The latter $z$'s match the redshift range of most of giant radio relics; for instance, CIZA J2242.8+5301 and RX J0603.3+4214 have $z=0.188$ and 0.225, respectively (see the references in the Introduction). Finally, for the uniformity of the sample, we chose clusters with the X-ray weighted temperature $T_{\rm X} \sim 5$ keV after merger. CIZA J2242.8+5301 and RX J0603.3+4214, on the other hand, are observed to have $T_{\rm X} \sim 7 - 10$ keV \citep[see, e.g.,][]{ogrean2013b, akamatsu15}, higher than those of sample clusters. Even with $\sigma_8=1.05$, the box size of our simulations is not large enough to produce such massive clusters. In the end, five sample clusters were complied from ten simulations with different initialization, and their characteristic parameters are listed in Table \ref{tab:t1}. The average virial masses of merging clumps (baryons plus DM) are $\left< M_{\rm heavy}\right> \sim 3.33 \times 10^{14} M_{\odot}$ and $\left< M_{\rm light} \right> \sim 1.67 \times 10^{14} M_{\odot}$, respectively, and the average mass of clusters after merger is $\sim 5 \times 10^{14} M_{\odot}$.}

{  Our simulated clusters have a resolution lower than those typically generated using either smoothed particle hydrodynamic (SPH) or adaptive mesh refinement (AMR) codes. So very weak shocks in the core region may not be fully resolved. But we here mainly concern shocks in the outskirts, which are observed as X-ray shocks and radio relics (see the Introduction). The statistics of those shocks reasonably converge at this resolution \citep[see][]{hong2014}, and also agree those of SPH and other codes \citep[see][]{vazza2011b}. Also our clusters were generated without including radiative and feedback processes. While these effects would be important in the core region, shocks in the cluster outskirts are less affected by such nongravitational effects, as shown in, for instance, \citet{kang2007}.} 

\subsection{Shock Identification and Energy Flux Calculation}
\label{sec:s2.2}

{  In sample clusters, shocks (actually grid zones containing shocks) were identified with the algorithm described in \citet{ryu2003} and \citet{hong2014} (See \citet{vazza2011b} for comparisons of different shock identification algorithms). Shocked grid zones were tagged if they satisfy the following three conditions: (1) $\nabla \cdot v < 0$, i.e., converging local flow, (2) $\Delta T \times \Delta \rho > 0$, i.e., same temperature and density gradient signs, and (3) $\left|\Delta \log T\right| > 0.11$, i.e., the Mach number greater than 1.3. In numerical simulations, shocks are represented by jumps typically spread over 2 - 3 zones, and ``shock zones'' were defined as the minima of $\nabla \cdot v$.} The sonic Mach number can be estimated from the temperature jump across the shock jump, $T_2/T_1=(5M_s^{2}-1)(M_s^{2}+3)/(16M_s^{2})$. Here, the subscripts 1 and 2 denote the preshock and postshock quantities, respectively. {  The Mach number of shock zones was defined as $M_s =$ max$(M_{s,x}, M_{s,y}, M_{s,z})$.} Very weak shocks are not energetically important, yet are easily confused with sonic waves, so only shocks with $M_s \geq 1.5$ were considered. {  Note that a shock surface consists of a number of shock zones.}

At shock zones, the shock kinetic energy flux was calculated as
\begin{equation}
f_{\phi} = (1/2) \rho_1 v_s^{3},
\end{equation}
where $v_s = M_s(\gamma P_{th,1}/\rho_1)^{1/2}$. A part of the shock kinetic energy is dissipated to accelerate CRs via DSA as well as to heat the gas, since ICM shocks are collisionless, as noted in the Introduction. The energy flux of CRs emerging from shock zones was estimated as
\begin{equation}
f_{\rm CR}=\eta(M_s) \times f_{\phi}
\end{equation}
\citep[see, e.g.,][]{ryu2003}. {  Here, $\eta(M_s)$ is the CR acceleration efficiency as a function of Mach number, and we employed the model presented in \citet{kangryu2013}. While our model $\eta$ converges to $\sim 0.23$ for strong shocks with $M_s \gg 1$, it is much smaller with 
$7\times 10^{-3} \la \eta \la 4\times 10^{-2}$ for $3\la M_s \la 4$, and almost negligible for $M_s \le 2$ (see Figure 2 of \citet{hong2014}). 
Such behavior of $\eta$ is consistent with the recent hybrid plasma simulations by \citet{caprioli2014}, although the magnitudes of two model $\eta$'s differ 
by up to a factor of two in the shock parameter range where a comparison can be made.
This difference is not important here, since we concern mainly the relative importance of shocks with different Mach numbers, rather than the absolute amount of CR generation at these shocks.}

The integrated kinetic and CR energies through shock surfaces were also calculated as
\begin{equation}
F_{\phi{\rm \ or \ CR}} = \sum_{\rm shocks} f_{\phi{\rm \ or \ CR}}\ \Delta S,
\label{eqintflu}
\end{equation}
where $\Delta S$ is the surface area of shock zone. 

In the following section, we will present the quantitative properties of merger shocks averaged over the entire population in the our sample clusters of relatively uniform characteristics.

\section{Results}
\label{sec:s3}

\begin{figure}[t]
\vskip 0.2cm
\hskip 0 cm
\centerline{\includegraphics[width=0.48\textwidth]{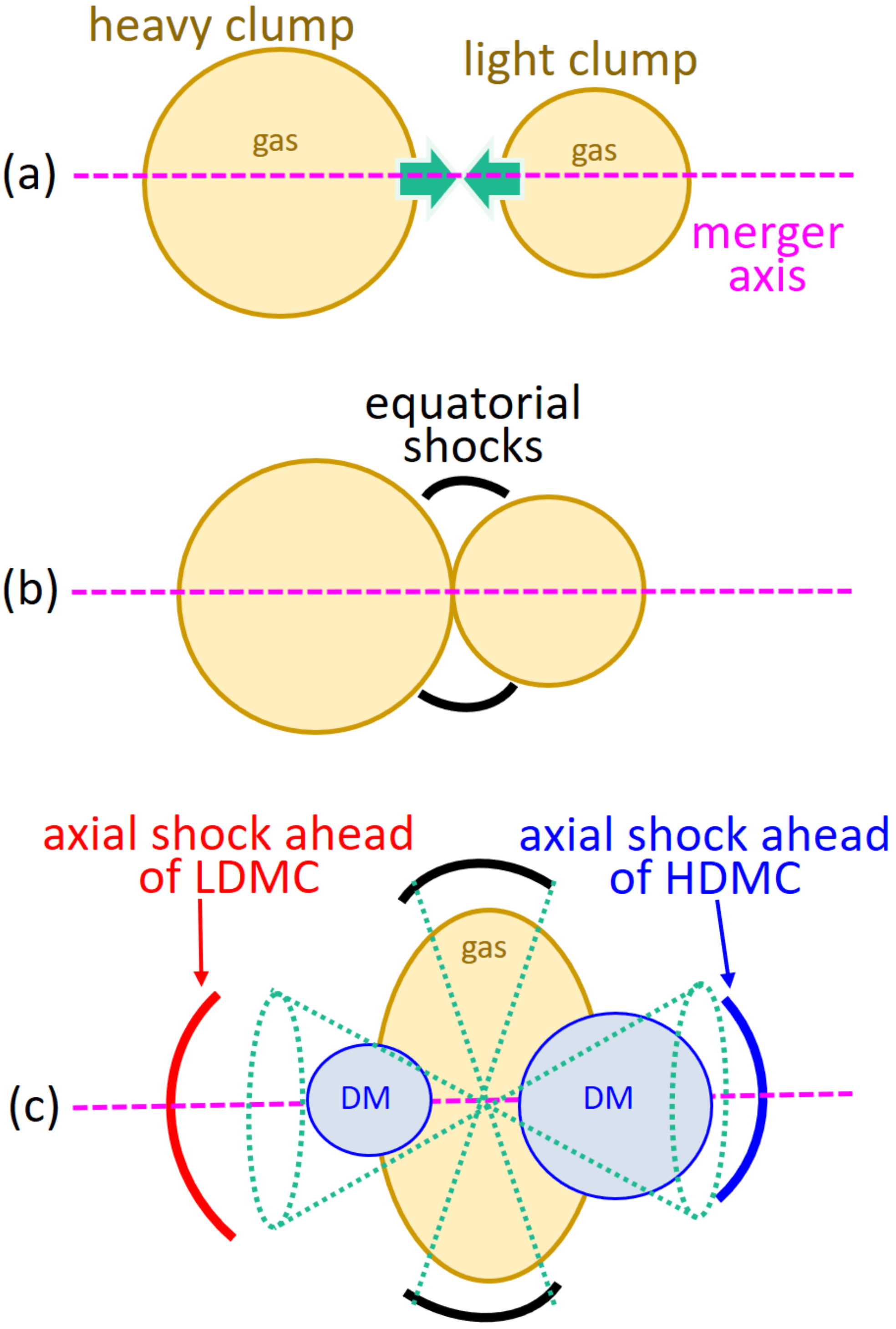}}
\vskip 0.1cm
\caption{Schematic picture of an idealized binary merger. (a) Heavy and light gas clumps undergo a head-on approach. (b) As the gas is compressed along the merger axis, ``equatorial shocks'' first expand outwards in the equatorial plane, perpendicular to the merger axis. (c) Later, ``axial shocks'' launch in the opposite directions along the merger axis and a single gas core forms. Light DM clump (LDMC) and heavy DM clump (HDMC) are also shown. Green lines in (c) draw the cone and disk over which the average properties of axial and equatorial shocks are calculated (see Section \ref{sec:s3.2}). \label{fig:f1}}
\end{figure}

\begin{figure*}[t]
\centering
\vskip 0.2cm
\centerline{\includegraphics[width=0.82\linewidth]{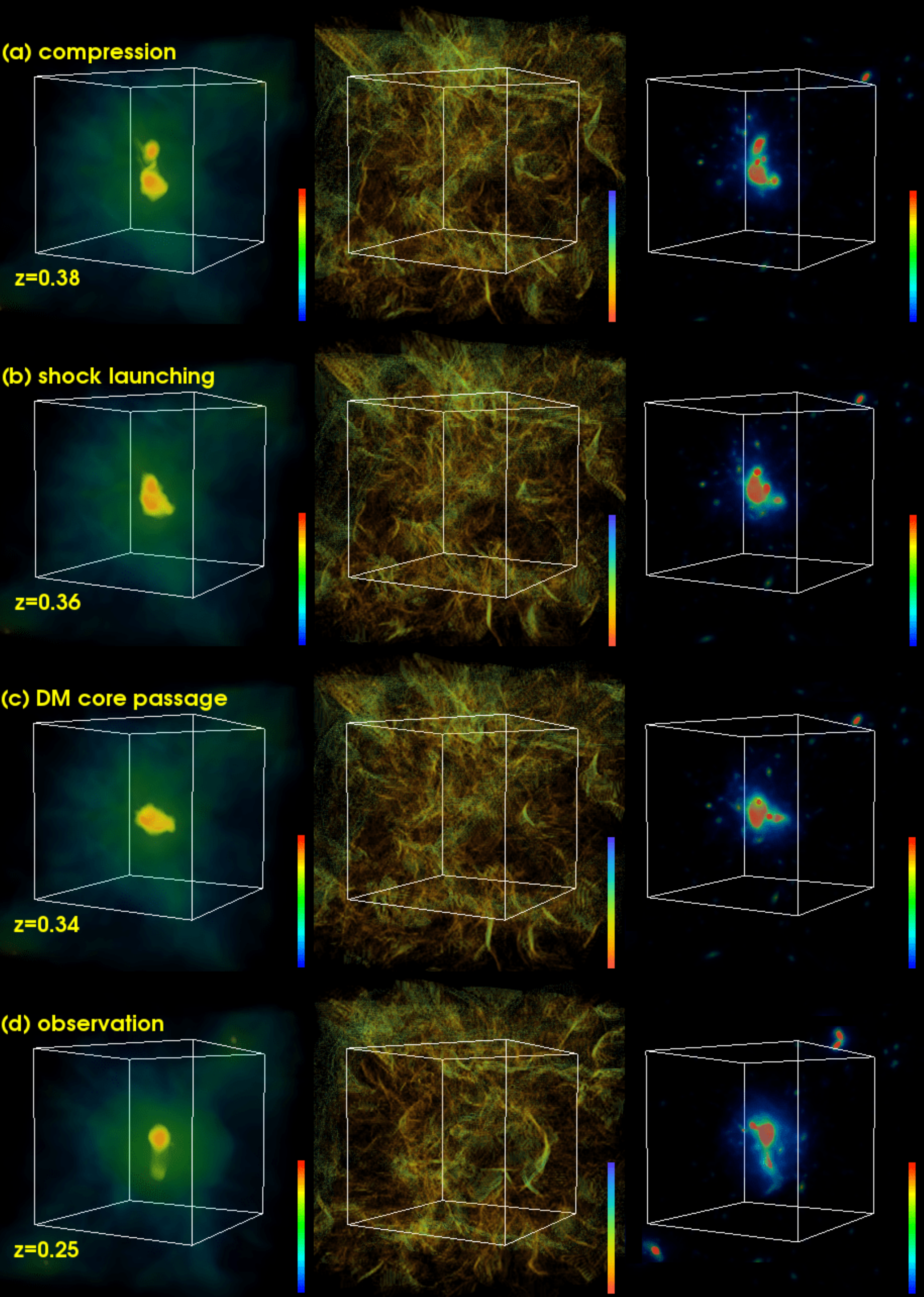}}
\caption{Cluster 1 of Table \ref{tab:t1} at four different epochs: (a) compression, (b) shock launching, (c) DM core passage and single gas core formation, and (d) the time of radio relic observation. A comoving box of 5.7 Mpc size is shown in white. Left panels show the X-ray emissivity in a logarithmic scale spanning $L_x = 10^{48}$ ergs/s (red) to $10^{39}$ ergs/s (blue), middle panels show the shock Mach numbers in a linear scale spanning $M_s = 7$ (blue) to 1.5 (red), and right panels show the DM density in a logarithmic scale spanning $\rho_{\rm DM}/\left< \rho_{\rm DM} \right> = 2 \times 10^3$ (red) to 1 (blue). \label{fig:f2}}
\end{figure*}

\begin{figure*}[t]
\centering
\vskip -1.3cm
\hskip -1.7cm
\centerline{\includegraphics[width=1.2\textwidth]{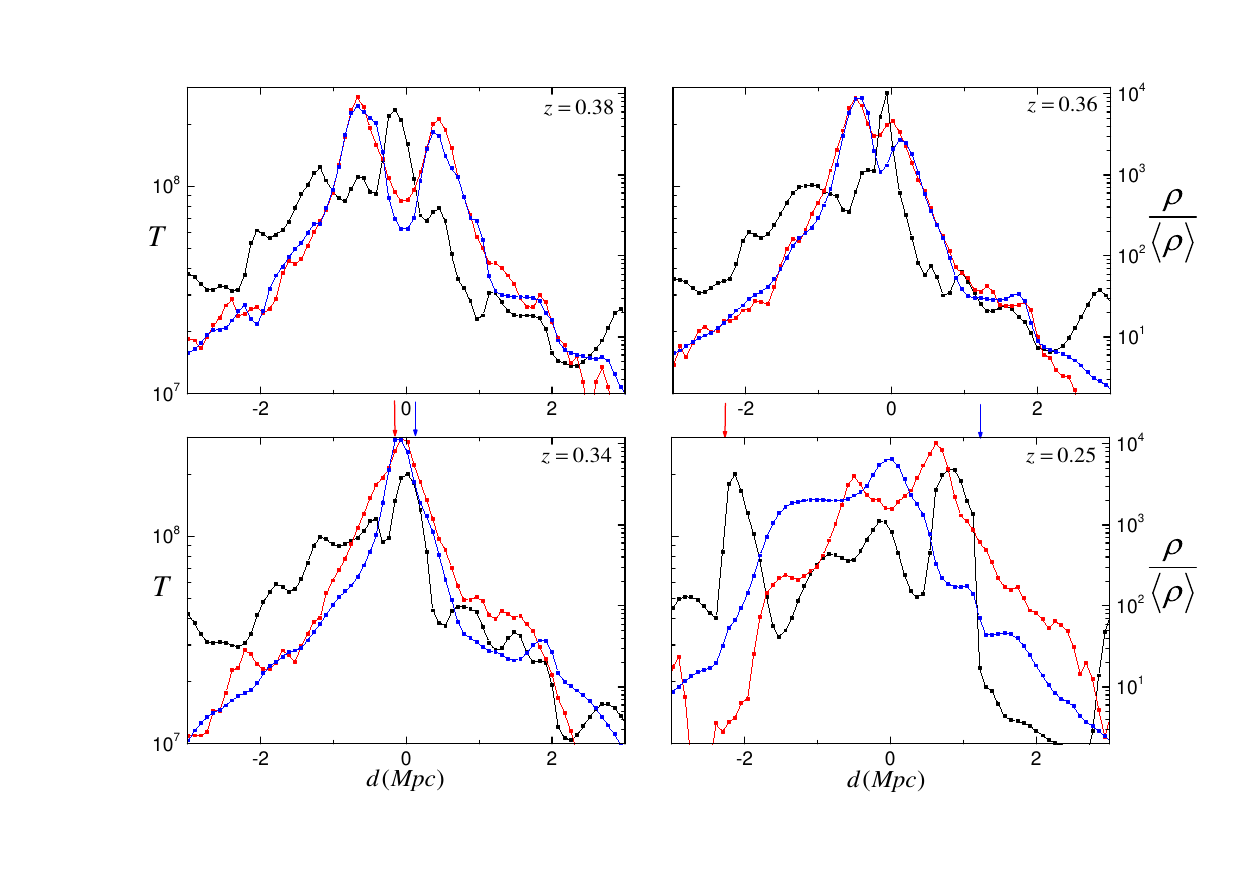}}
\vskip -1.4cm
\caption{1D distributions along the merger axis in Cluster 1 of Table \ref{tab:t1} at four epochs, same as those in Figure \ref{fig:f2}. The gas temperature (in units of K, black), the gas density ($\rho_{\rm gas}/\left< \rho_{\rm gas} \right>$, blue), and the DM density ($\rho_{\rm DM}/\left< \rho_{\rm DM} \right>$, red) are shown. In the panels of $z = 0.34$ and 0.25, the axial shock ahead of LDMC is marked with red arrows, while the axial shock ahead of HDMC is marked with blue arrows. \label{fig:f3}}
\end{figure*}

\begin{figure*}[t]
\centering
\vskip 0.1cm
\centerline{\includegraphics[width=1\textwidth]{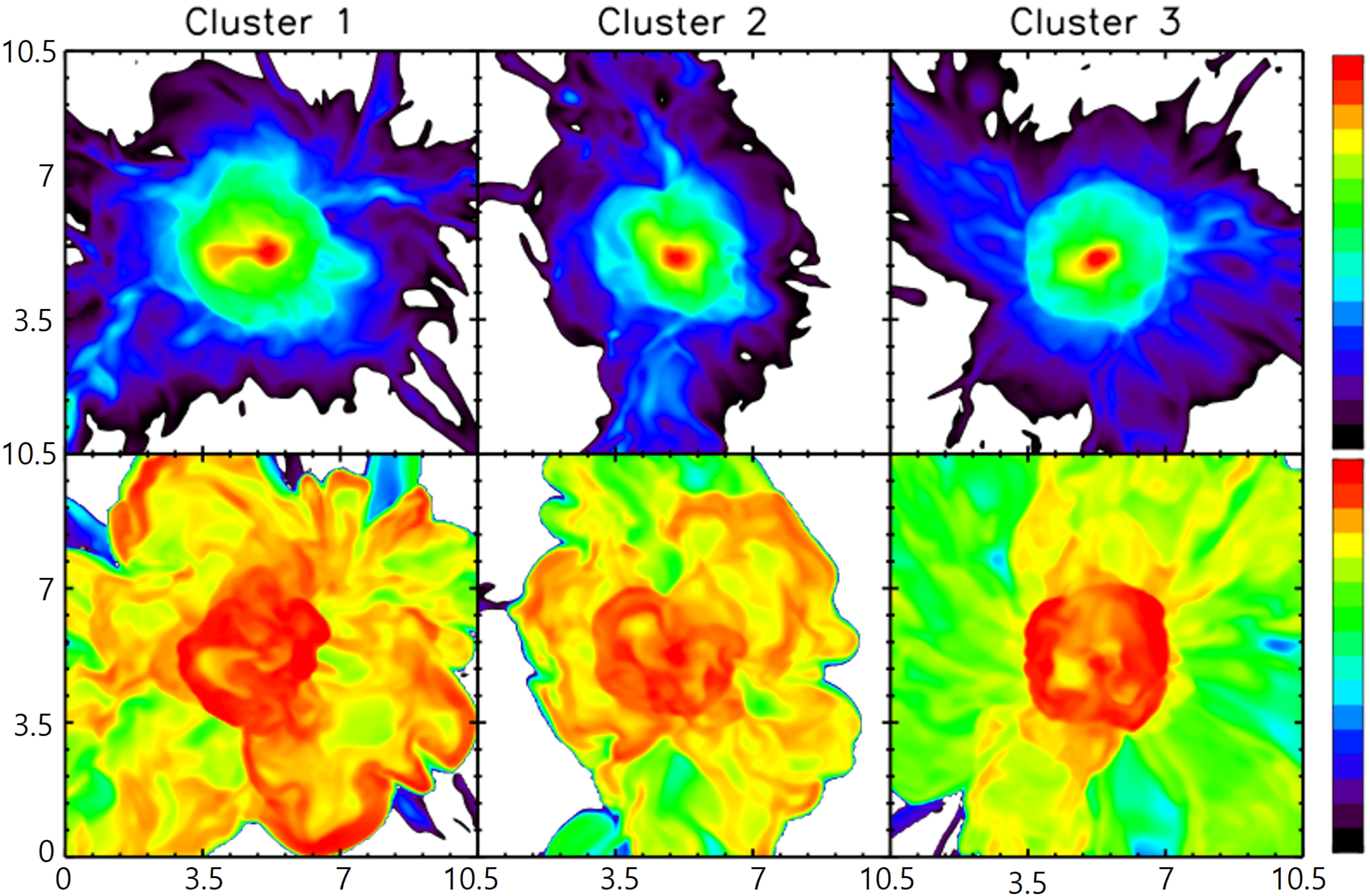}}
\vskip 0cm
\caption{2D slices of area (10.5 Mpc)$^2$ through X-ray peaks at $z = 0.25$, $0.24$, and $0.14$ (the times of radio relic observation) for Clusters 1, 2, and 3 of Table \ref{tab:t1}, respectively.  Here, the X-ray peak is at the center of each image. Top panels show the gas density in a logarithmic scale spanning $\rho_{\rm gas}/\left< \rho_{\rm gas} \right> = 10^4$ (red) to 1 (blue) and bottom panels show the gas temperature in a logarithmic scale spanning $T = 10^8$ K (red) to $10^4$ K (blue). \label{fig:f4}}
\end{figure*}

\begin{figure*}[t]
\vskip -1cm
\hskip 0.1cm
\centerline{\includegraphics[width=1.22\textwidth]{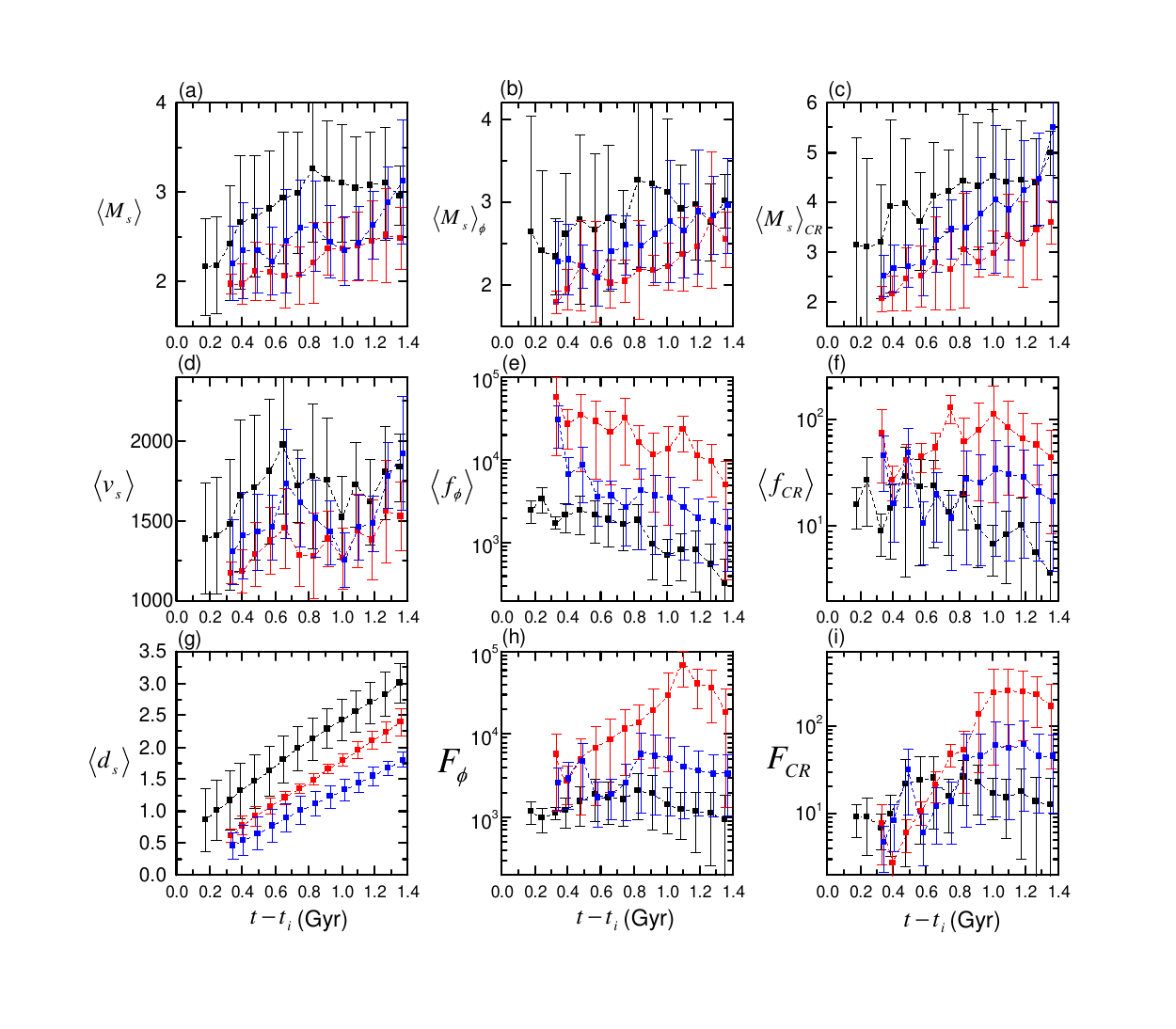}}
\vskip -1.7cm
\caption{Time evolution of merger shock properties during $\sim 1.4$ Gyr after the axial shock launching time, $t_i$. Red is for axial shocks ahead of LDMCs, blue is for axial shocks ahead of HDMCs, and black is for equatorial shocks. (a) Mach number $\left< M_s \right>$, (b) $f_{\phi}$-weighted Mach number $\left< M_s \right>_{\phi}$, (c) $f_{\rm CR}$-weighted Mach number $\left< M_s \right>_{\rm CR}$, (d) shock speed $\left< v_s \right>$, (e) kinetic energy flux through shock surfaces $\left< f_{\phi} \right>$, (f) CR energy flux produced at shocks $\left< f_{\rm CR}\right>$, (g) proper distance from the cluster center $\left< d_s \right>$, (h) integrated kinetic energy through shock surfaces $F_{\phi}$, and (i) integrated CR energy produced at shocks $F_{\rm CR}$ are shown. Here, $v_s$ is in units of km s$^{-1}$, $d_s$ in units of Mpc, $f$'s in units of $10^{40}$ ergs s$^{-1}$ Mpc$^{-2}$, and $F$'s in units of $10^{40}$ ergs s$^{-1}$. Squares and error bars denote averages and 1 $\sigma$ deviations. Except $F_{\phi}$ and $F_{\rm CR}$, the averages and standard deviations were calculated for all shock zones found in five sample clusters. For $F_{\phi}$ and $F_{\rm CR}$, the averages and standard deviations are for five sample clusters. \label{fig:f5}}
\end{figure*}

\subsection{Overview of Merging Process}
\label{sec:s3.1}

To set the stage for describing the merging process, we begin with a general overview of an idealized binary merger \citep[e.g.][]{vanweeren2011a, molnar2017a}, as illustrated in Figure \ref{fig:f1}. As two clumps are approaching and being compressed, shocks form and first move outwards in the equatorial plane, perpendicular to the merger axis. We name these shocks {\it equatorial shocks}.  Later, two {\it axial shocks} launch into the opposite directions along the merger axis. The {\it core passage} of DM clumps and the formation of a single gas core occur after the shock launch.

Mergers in our structure formation simulations are, of course, much more complex. They excite turbulent flow motions and are often accompanied by multiple minor mergers and secondary infall along connecting filaments. {  As a consequence, the formation of merger shocks proceeds in a way more complex than in idealized binary mergers.}

Figure \ref{fig:f2} shows the merging process in a representative cluster, Cluster 1 (see Table \ref{tab:t1}). Two clumps, composed of baryons and DM, are approaching in an almost head-on collision, and for the sake of convenience, we refer to the four epochs in Figure \ref{fig:f2} as the following terms:
(a) compression phase ($z=0.38$) during which the two clumps are approaching,
(2) shock launching phase ($z=0.36$) when the first axial shocks launch,
(3) DM core passage phase ($z=0.34$) when the two DM cores pass each other and two gas clumps merge to form a single core, 
(4) the time of radio relic observation ($z=0.25$) at $\sim 1$ Gyr after the first axial shocks launch (see Section \ref{sec:s3.2}).

The middle column of Figure \ref{fig:f2} demonstrates the presence of complex networks of shock surfaces in the ICM, even before the two clumps begin to contact and get compressed. Shock surfaces formed during the merger are patchy and highly intermittent with filamentary structures of high Mach number regions.

Figure \ref{fig:f3} shows the one-dimensional (1D) distributions of the gas temperature (black), gas density (blue), and DM density (red) along the merger axis of Cluster 1 at the same four epochs as those of Figure \ref{fig:f2}. We define the zero point of distance, $d$, along the merger axis as the position of maximum X-ray peak at a given time, except at earlier epochs ($z=0.38$ and $0.36$); at $z = 0.36$, $d = 0$ corresponds to the X-ray peak appeared during the compression, and at $z=0.38$, the same zero point as at $z=0.36$ is adopted. The panel for $z=0.38$ shows that the heavy (light) gas and DM clumps are approaching from the left-hand side (right-hand side). The gas clumps are being compressed, and a temperature peak appears between them. In the panel for $z=0.36$, while the density peaks of gas clumps are still getting close to each other, axial shocks start to form at $d=0$. The panel for $z=0.34$ shows a single DM peak around the zero point, indicating it is close to the DM core passage epoch. At that time, both the gas density and temperature distributions have a single peak at $d=0$, telling the formation of a merged core. Two axial shocks in the both sides of the peak are apparent.

The last epoch at $z=0.25$ represents the time, around when the axial shocks have the ``best chance'' to be observed as a radio relic or double radio relics (see Section \ref{sec:s3.2}). This corresponds to $\sim 1$ Gyr after the axial shock launching. The axial shocks are identified at $d_s \sim 1 - 2$ Mpc from the X-ray peak. The Mach number of the axial shock traveling ahead of the light DM clump (LDMC, hereafter) is $M_s \approx 3.5$ (the red arrow), while that of the shock  traveling ahead of the heavy DM clump (HDMC, hereafter) is $M_s \approx 4$ (the blue arrow). Earlier studies for idealized binary mergers also found that the shock ahead of HDMC is stronger than that ahead of LDMC \citep{vanweeren2011a, molnar2017a}.

Figure \ref{fig:f4} shows the two-dimensional (2D) slices of the gas density (top panels) and temperature (bottom panels), passing through X-ray peaks in the first three clusters of Table \ref{tab:t1} at the times of radio relic observation. In all three sample clusters, the heavy clumps approached from the left-hand side, while the light clumps came from the right-hand side. The clumps merged into cores, and the X-ray peaks are located at the center of the images. The elongation axes of density cores roughly represent the merger axes. Merger shocks (both axial shocks and equatorial shocks) are manifested as sharp jumps in the temperature and density distributions. In addition, structures induced by turbulent flow motions and other dynamic activities are evident.

\subsection{Properties of Merger Shocks}
\label{sec:s3.2}

\begin{figure*}[t]
\centering
\vskip -0.2cm
\hskip -1.5cm
\centerline{\includegraphics[width=1.15\textwidth]{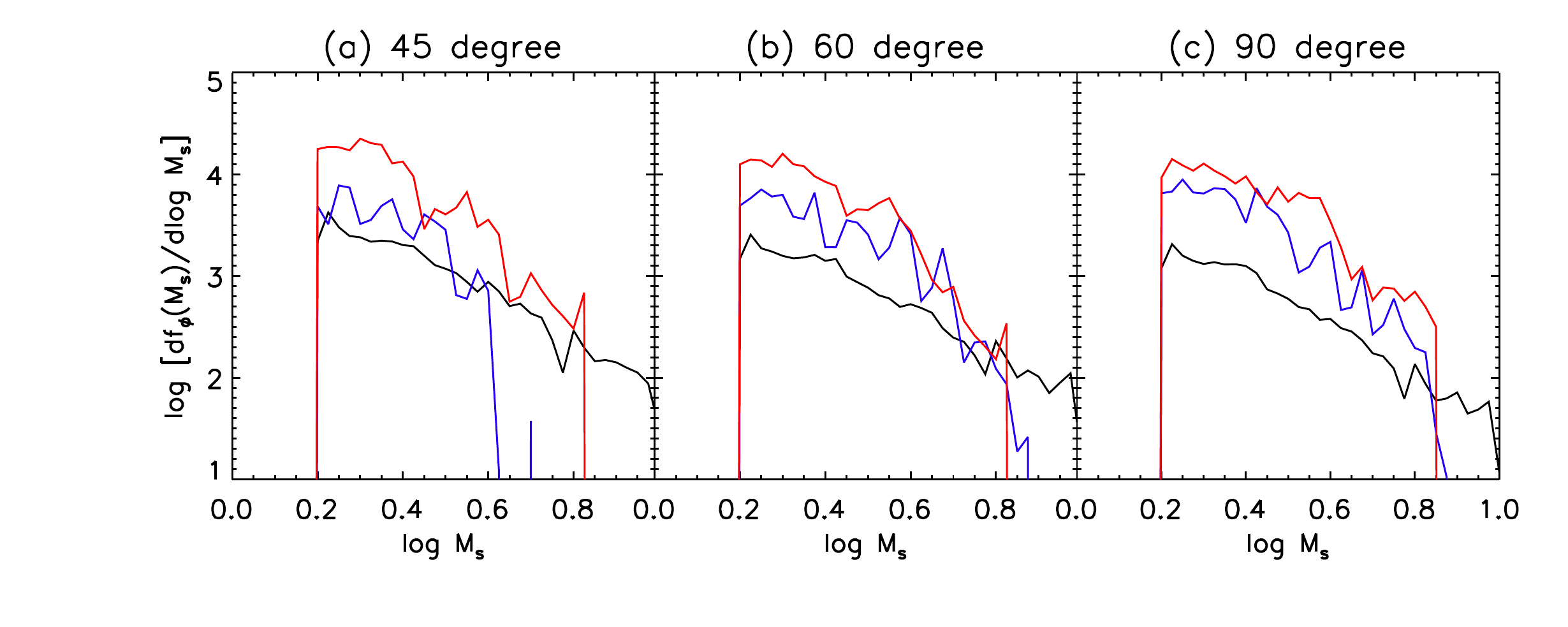}}
\vskip -0.8cm
\caption{{  Differential shock kinetic energy flux as a function of Mach number for different opening angles, $\Delta \theta$'s, for Cluster 1 at $z = 0.25$ (the time of radio relic observation). Here, $f_{\phi}$ is in units of $10^{40}$ ergs s$^{-1}$ Mpc$^{-2}$. Red is for axial shocks ahead of LDMC, blue is for axial shocks ahead of HDMC, and black is for equatorial shocks. \label{fig:f6}}}
\end{figure*}

In our sample merging clusters from LSS formation simulations, in addition to merger-driven shocks, numerous shocks form and disappear as a consequence of background activities. So it is not straightforward to clearly isolate merger shocks from shocks of other types (i.e., turbulent and infall shocks) and track their evolution. Bearing the merging process described above in mind, we attempted to pick up merger shocks ``visually'' using three-dimensional (3D) images like in those in Figure \ref{fig:f2} as well as 2D slices and 1D line-cut plots as those in Figures \ref{fig:f4} and \ref{fig:f3}.

Merger shocks were divided into three different categories, axial shocks ahead of LDMC, axial shocks ahead of HDMC, and equatorial shocks. As illustrated in Figure \ref{fig:f1}, for the axial shocks we counted those within the two polar cones with opening angles of $\Delta \theta = 45^{\circ}$, confined by either the polar angle of $\theta\le 22.5^{\circ}$ or $\theta\ge 157.5^{\circ}$ around the merger axis. For the equatorial shocks, we considered those located within a disk-like zone confined by $-22.5^{\circ} \le (\theta - 90 ^{\circ}) \le 22.5^{\circ}$ around the equatorial plane. Here, the cluster center is the peak of X-ray emission at each epoch. The mean distance, $\left< d_s \right>$, of the three shock categories were estimated by taking the average value of $d_s$ of the shocks that belong to each category. Then, the shock selection process was repeated and refined with shocks within $\left< d_s \right> \pm 0.3$ Mpc for axial shocks, and with shocks within $\left< d_s \right> \pm 0.5$ Mpc for equatorial shocks. A larger distance span was considered for equatorial shocks, because they form over $360^\circ$ of the azimuthal angle, thus showing a larger fluctuation in the position. Figure \ref{fig:f5}(g) shows the time evolution of the converged value of $\left< d_s \right>$. In this iteration procedure, some of merger shocks could be missed, and some shocks of other types, particularly turbulent shocks, could be counted in erroneously. But we find that the statistical properties of merger shocks are overall not very sensitive to the choice of the opening angle, nor to the distance spanning. {  Figure \ref{fig:f6}, for instance, shows the kinetic energy flux through shock surfaces, $f_\phi$, a statistics presented below (see Figure \ref{fig:f5}(e)), for different $\Delta \theta$'s, demonstrating its insensitivity to $\Delta \theta$.}

{  As a reference point of time, we use the {\it axial shock launching time}, $t_i$, which was calculated as follows. Once two type of axial shocks were identified, the average distance between them as a function of time, $D(t)$, was estimated. At the initial stage of mergers, shocks are hard to be identified reliably, since they are very weak with $M_s \sim 1 - 2$ and form in the turbulent core regions. Reliable identification of axial shocks becomes feasible typically after the core passage epoch. So $t_i$ was calculated by extrapolating the shock distance backward in time, that is, as the time when $D(t_i) = 0$. The redshift of $t_i$ is given in the last column of Table \ref{tab:t1}.}

{  The selection of merger shocks and the calculation of their statistics were made at 15 epochs after $t_i$ with separation $\Delta z \sim 0.01$ (corresponding $\Delta t \sim 0.09$ Gyr) for the five sample clusters in Table 1.} The means and dispersions of shock properties were calculated for each shock category over all the shocks detected in five sample clusters. The number of counted shocks (shock zones) in each category increases from $N_s \lesssim 100$ at the axial shock launching time to $\sim 1500 - 2000$ (corresponding to the shock surface area of $\sim 2 - 3$ (Mpc)$^2$) during the time period of 1.4 Gyr in each sample cluster.

In Figure \ref{fig:f5}, the mean properties of axial shocks ahead of LDMC (red), axial shocks ahead of HDMC (blue), and equatorial shocks (black) are presented as a function of time, counted from $t_i$. Although the statistical fluctuations (error bars in the figure) are rather large, the mean values exhibit clear trends. { We note that the noisiness of these physical quantities should come from the inherent nature of merger shocks induced in the turbulent ICM during the hierarchical structure formation.}

As shown in Figure \ref{fig:f5}(g), the mean distance of equatorial shocks is the largest among three shock categories, since they launch earlier. And $\left< d_s \right>$ for axial shocks ahead of LDMC is larger than that for axial shocks ahead of HDMC, indicating that the X-ray peaks are close to HDMCs in our merging clusters. The mean distances of all three shock categories increase in time, and axial shocks, for instance, reach $\left< d_s \right> \sim 1-2$ Mpc by the time $t - t_i \simeq 1$ Gyr.

The top panels of Figure \ref{fig:f5} show the mean values of shock Mach numbers, $\left< M_s \right>$, $\left< M_s \right>_{\phi}$ weighted with shock kinetic energy flux ($f_{\phi}$), and $\left< M_s \right>_{\rm CR}$ weighted with CR energy flux ($f_{\rm CR}$). Firstly, the Mach numbers overall increase in time, while the mean shock speed, $\left< v_s \right>$, in Figure \ref{fig:f5}(d) increases in the early phase during $\sim 0.6$ Gyr, but then fluctuates afterward. The overall increase of shock Mach numbers in the late stage reflects the fact that the gas temperature tends to decrease in the cluster outskirts ($\gtrsim 1$ Mpc), as can be seen in Figures \ref{fig:f3} and \ref{fig:f4}. Secondly, both $\left< v_s \right>$ and $\left< M_s \right>$ of equatorial shocks have the largest values, since they propagate mostly to low density regions surrounding merging clumps. Moreover, $\left< v_s \right>$ and $\left< M_s \right>$ of axial shocks ahead of HDMC are larger than those of axial shocks ahead of LDMC. Yet, $\left< d_s \right>$ for axial shocks ahead of HDMC is smaller, owing to the fact that $\left< d_s \right>$ includes not only the propagation of shocks but also the displacement of X-ray peaks. Thirdly, while $\left< M_s \right>$ and $\left< M_s \right>_{\phi}$ are comparable, $\left< M_s \right>_{\rm CR}$ is larger by about unity or so, especially in the late stage. This is due to the dependence of the CR acceleration efficiency, $\eta(M_s)$, on the shock Mach number; our model $\eta(M_s)$ is larger for stronger shocks (see Section \ref{sec:s2.2}).

The average kinetic energy flux through shock surfaces, $\left< f_{\phi} \right>$, in Figure \ref{fig:f5}(e) tends to decrease in time as shocks move outwards, since the gas density decreases in the cluster outskirts. {  The average CR energy flux produced at shocks, $\left< f_{\rm CR} \right>$, in Figure \ref{fig:f5}(f), on the other hand, shows complicated time evolution, reflecting the Mach number dependence of the CR acceleration efficiency.} Both $\left< f_{\phi} \right>$ and $\left< f_{\rm CR} \right>$ for axial shocks ahead of LDMC are the largest, because these shocks propagate into the gas with higher density that is originally associated with heavier gas clumps.

Figures \ref{fig:f5}(h) and \ref{fig:f5}(i) show the shock kinetic and CR energy fluxes integrated over shock surfaces, $\left< F_{\phi} \right>$ and $\left<F_{\rm CR} \right>$, averaged for five sample clusters. Again, $\left< F_{\phi} \right>$ and $\left< F_{\rm CR} \right>$ of axial shocks ahead of LDMC are the largest, while those of equatorial shocks are the smallest. Although axial shocks ahead of LDMC are the weakest with smallest $\left< M_s\right>$ among shocks of three categories, they are energetically the most important; that is, they process the largest amount of kinetic energy and also generate the largest amount of CRs, especially at late times of $t - t_i \simeq 0.8 - 1.4$ Gyr.

In particular, in Figure \ref{fig:f5}(h), $F_{\phi}$ for axial shocks ahead of LDMC peaks at $\sim 10^{44} - 10^{45}$ ergs s$^{-1}$ during $t-t_i \simeq 0.8 - 1.4$ Gyr. The total energy processed during the period is $\sim$ several $\times 10^{60}$ ergs, which is a substantial fraction of the merger energy $\sim 10^{62}$ ergs (see the Introduction). $F_{\phi}$ for axial shocks ahead of HDMC is about an order of magnitude smaller, and $F_{\phi}$ for equatorial shocks is even smaller by a factor of several. So axial shocks ahead of LDMC should have the best chance to be observed as X-ray shocks, especially at $\sim 1$ Gyr after the launching of the shocks. During the peak period, for these axial shocks, the $f_{\phi}$-weighed Mach number ranges $\left< M_s \right>_{\phi} \sim 2 - 3$ and their distance from the cluster center ranges $\left< d_s \right> \simeq 1 - 2$ Mpc. These are in reasonable agreement with the observed characteristics of X-ray shocks, as noted in the Introduction.

In Figure \ref{fig:f5}(i), again, $F_{\rm CR}$ for axial shocks ahead of LDMC is several to tens times larger than $F_{\rm CR}$ for other category shocks during $t-t_i \simeq 0.8 - 1.4$ Gyr. So they should have the best chance to light up as radio relics and thus be observed in radio. The $f_{\rm CR}$-weighed Mach number for the shocks during the peak period is, on the other hand, in the range of $\left< M_s \right>_{\rm CR} \sim 3 - 4$, higher than $\left< M_s \right>_{\phi}$, as noted above. This range of $\left< M_s \right>_{\rm CR}$ is consistent with the range of the shock Mach numbers estimated from the radio spectral indices of observed radio relics (see the references in the Introduction). The potential manifestation of larger $M_s$ in radio relic observations than in X-ray shock observations was pointed out in \citet{hong15}. Our results confirm such tendency, indicating that the difference between $M_{\rm X}$ and $M_{\rm radio}$ might be due to the representations of different parts of shock surfaces, that is, higher $M_s$ parts for radio observations while lower $M_s$ parts for X-ray observations, as noted in the Introduction. The range of $\left< d_s \right> \simeq 1 - 2$ Mpc for axial shocks ahead of LDMC during the peak of CR production is also comparable to the positions of observed radio relics.

{  We note that the Sunyaev-Zel'dovich (SZ) decrements found CIZA J2242.8+5301 could be interpreted as high pressure regions generated by equatorial shocks propagating in the direction perpendicular to the merger axis \citep[see][]{rumsey2017}. An X-ray temperature break found in the merger system between Abell 399 and Abell 401 could also indicate a signature of equatorial shocks \citep[see][]{akamatsu17}. Although the nature of these observed features should be further investigated, their positions are consistent with the equatorial shocks defined in this study. These indicate that although energetically sub-dominant, equatorial shocks could have possibly observable imprints.}

From Figure \ref{fig:f5}(d), one can see that $\left< v_s \right>$ increases during the peak period of $F_{\phi}$ and $F_{\rm CR}$. The adiabatic blast wave solution for a point explosion requires the density gradient steeper than $\rho^{-3}$ for accelerating shock fronts \citep[see, e.g.,][]{ryu1991}. Apparently, the blast wave assumption does not hold for merger-driven shocks, since the kinetic and gravitational energies of merging clumps are continuously dissipated through shocks and additional energies are supplied by secondary infall and multiple minor mergers. In addition, $\left< v_s \right>$ includes the contributions not only from the shock propagation, but also from turbulent flow motions ahead of shocks. As a matter of fact, large fluctuations in $\left< v_s \right>$ as well as in $\left< M_s \right>$ reflect complicated flow dynamics of clusters.

\subsection{Dark Matter Distribution}
\label{sec:s3.3}

From Figures \ref{fig:f2}, \ref{fig:f3} and \ref{fig:f4}, we can see the relative positions of merger shocks, X-ray peak, and DM clumps in Cluster 1 at $z=0.25$ or $t-t_i \simeq 1$ Gyr, an epoch close to the peak of $F_{\phi}$ and $F_{\rm CR}$. Particularly, in Figure \ref{fig:f3}, with the peak of X-ray emission located in the middle ($d = 0$), the axial shock ahead of LDMC is at $d\approx -2.3$ Mpc, whereas the HDMC is at $d\approx 0.6$ Mpc. Shocks ahead of LDMC are the most energetic, as discussed above. Assuming that the shock appears as the main radio relic, this configuration is consistent with that of the Sausage relic in CIZA J2242.8+5301, at least qualitatively \citep[see][]{akamatsu15, okabe2015}. In RX J0603.3+4214, on the other hand, the Toothbrush relic is located close to the HDMC, as mentioned in the Introduction. While our sample clusters undergo almost head-on collisions of mass ratio $\sim 2$ clumps, the Toothbrush relic seems to have been produced by a merger involving multiple clumps \citep[see][]{brug12b, jee2016}.  The detailed distributions of shocks, X-ray emission, and DM in merging clusters should depend on a number of parameters, including the number of clumps and their masses and impact parameters. Studies of the dependences on such parameters would need a much large sample of merging clusters and are beyond the scope of this paper.

\section{Summary}
\label{sec:s4}

In the currently favored paradigm of hierarchical structure formation, galaxy clusters form through successive mergers of sub-cluster clumps, and shock waves are naturally induced as a consequence. Major mergers in relatively recent epochs of $z < 0.5$ are among the most energetic events in the universe, and the merger shocks associated with them are observed in X-ray and radio. 

{  In this study, we examined the properties of merger shocks in galaxy clusters from cosmological hydrodynamic simulations for the LSS formation of the universe. We first compiled a sample of five merging clusters in ten simulations with different initialization; all undergo through almost head-on collisions of mass ratio $\sim 2$ at $z < 0.5$, which result in merged systems with $T_{\rm X} \sim 5$ keV. We then isolated shocks produced by merger activities, and quantified their properties such as the shock speed, Mach number, and shock energy flux. 
Due to the turbulent nature of the ICM, the properties of the shocks can be described only statistically with means and standard deviations for a population of identified shocks associated with merger events. We also calculated the time evolution of those shock properties.}

We described the merging process in our sample clusters as follows (see Figures \ref{fig:f2} and \ref{fig:f3}).
(a) As the gas is compressed during the approach of two clumps, ``equatorial shocks'' launch near the equatorial plane toward the direction perpendicular to the merger axis.
(b) As the clumps get closer, ``axial shocks'' launch along the merger axis. 
(c) The core passage of DM clumps and the formation of a single gas core occur after the launching of the axial shocks.
(d) X-ray shocks and radio relics are likely to be observed in the cluster outskirts ($\sim 1 - 2$ Mpc) at $\sim 1$ Gyr after the shock launching.

Our findings are summarized as follows.\hfill\break
(1) {  The surfaces of merger shocks are not smooth. The Mach number distribution on the surfaces is highly intermittent and the high Mach number parts form filamentary structures.}\hfill\break
(2) As the merger shocks propagate out from the cores to the outskirts, the shock Mach number, $M_s$, on average increases in time, while the shock speed does not necessarily.\hfill\break
(3) {  The kinetic energy flux through shock surfaces, $f_{\phi}$, decreases in time, since the gas density is lower in the outskirts. But the CR energy flux produced at shocks, $f_{\rm CR}$, shows complicated time evolution.}
\hfill\break
(4) Axial shocks propagating ahead of LDMC are most energetic. They process large amounts of the kinetic energy, $F_{\phi}$, and the CR energy, $F_{\rm CR}$, and thus have the best chance to be observed as X-ray shocks and radio relics.\hfill\break 
(5) $F_{\phi}$ and $F_{\rm CR}$ of axial shocks ahead of LDMC peak at $t - t_i \sim 1$ Gyr after the shocks launched, or when the shocks are located at $d_s \sim 1 - 2$ Mpc from the cluster center. At the time, the shocks have $\left< M_s \right>_{\phi} \simeq 2 - 3$ (weighted with $f_{\phi}$), while $\left< M_s \right>_{\rm CR} \simeq 3 -4$ (weighted with $f_{\rm CR}$). This is because the CR acceleration is more efficient at the parts of shock surfaces with higher Mach numbers.\hfill\break
(6) Both DM clumps survive through merger, and their peaks persist. In our sample clusters, after the DM core passage, the LDMC is located behind the most energetic axial shocks, while the HDMC lies in the other side but closer to the peak of X-ray emission which coincides with the gas core.

{  Finally, we note that the properties of merger shocks, as well as their positions relative to X-ray peak and DM clumps, should depend on a number of merger parameters. 
More comprehensive investigation of such dependence requires a very large sample of simulated merging clusters, and we will leave it as future works.}

\acknowledgments
{  We thank the anonymous referee for constructive comments.} J.-H. H. was supported by the National Research Foundation of Korea through grant 2017R1A2A1A05071429. D.R.was supported by the National Research Foundation of Korea through grant 2016R1A5A1013277.
H.K. was supported by the Basic Science Research Program of the NRF of Korea through grant 2017R1D1A1A09000567.

\end{document}